\documentclass[ reprint, aps, prl, two column, showkeys, superscriptaddress]{revtex4-1}
\usepackage{graphicx,epsfig}
\usepackage{amssymb}
\usepackage{amsmath}
\usepackage{bm}
\usepackage{textcomp}
\usepackage{soul,xcolor}

\usepackage{color}

\begin{document}
\setcounter{page}{1}
\setstcolor{red}

\title[]{Correlated states of 2D electrons near the Landau level filling $\nu=1/7$}
\author{Yoon Jang \surname{Chung}}
\affiliation{Department of Electrical Engineering, Princeton University, Princeton, NJ 08544, USA  }
\author{D. \surname{Graf}}
\affiliation{National High Magnetic Field Laboratory, Tallahassee, FL 32310, USA}
\author{L. W. \surname{Engel}}
\affiliation{National High Magnetic Field Laboratory, Tallahassee, FL 32310, USA}
\author{K. A. \surname{Villegas Rosales}}
\affiliation{Department of Electrical Engineering, Princeton University, Princeton, NJ 08544, USA  }
\author{P. T. \surname{Madathil}}
\affiliation{Department of Electrical Engineering, Princeton University, Princeton, NJ 08544, USA  }
\author{K. W. \surname{Baldwin}}
\affiliation{Department of Electrical Engineering, Princeton University, Princeton, NJ 08544, USA  }
\author{K. W. \surname{West}}
\affiliation{Department of Electrical Engineering, Princeton University, Princeton, NJ 08544, USA  }
\author{L. N. \surname{Pfeiffer}}
\affiliation{Department of Electrical Engineering, Princeton University, Princeton, NJ 08544, USA  }
\author{M. \surname{Shayegan}}
\affiliation{Department of Electrical Engineering, Princeton University, Princeton, NJ 08544, USA  }

\date{\today}

\begin{abstract}

The ground state of two-dimensional electron systems (2DESs) at low Landau level filling factors ($\nu\lesssim1/6$) has long been a topic of interest and controversy in condensed matter. Following the recent breakthrough in the quality of ultra-high-mobility GaAs 2DESs, we revisit this problem experimentally and investigate the impact of reduced disorder. In a GaAs 2DES sample with density $n=6.1\times10^{10}$ /cm$^2$ and mobility $\mu=25\times10^6$ cm$^2$/Vs, we find a deep minimum in the longitudinal magnetoresistance ($R_{xx}$) at $\nu=1/7$ when $T\simeq104$ mK. There is also a clear sign of a developing minimum in the $R_{xx}$ at $\nu=2/13$. While insulating phases are still predominant when $\nu\lesssim1/6$, these minima strongly suggest the existence of fractional quantum Hall states at filling factors that comply with the Jain sequence $\nu=p/(2mp\pm1)$ even in the very low Landau level filling limit. The magnetic field dependent activation energies deduced from the relation $R_{xx}\propto e^{E_A/2kT}$ corroborate this view, and imply the presence of pinned Wigner solid states when $\nu\neq p/(2mp\pm1)$. Similar results are seen in another sample with a lower density, further generalizing our observations.

%, suggesting the existence of a fractional quantum Hall state (FQHS

\end{abstract}
\maketitle

	High-quality, two-dimensional electron systems (2DESs) are versatile platforms that are often utilized to study many-body electron physics. At low enough temperatures, the kinetic energy of a 2DES is determined by the Fermi energy of the system, and its relative strength with respect to the Coulomb energy can be tuned by varying the electron density. Moreover, the application of a magnetic field perpendicular to the 2DES further enhances the influence of electron-electron interaction through Landau quantization of the density of states. Several exotic many-body electron phases have materialized in clean 2DESs using this framework, such as the odd- and even-denominator fractional quantum Hall states (FQHSs) \cite{Tsui,Willett}, Wigner solids \cite{Wigner1,Wigner2,Wigner3}, and nematic/stripe phases \cite{stripe1,stripe2}; for reviews, see \cite{ShayeganReview,JainBook,HalperinBook}.
	
	Many of the interaction-driven phenomena that emerge in experiments are understood fairly well \cite{ShayeganReview,JainBook,HalperinBook}. For example, theory using Laughlin's wavefunction \cite{Laughlin} or the composite fermion approach \cite{Jain} can successfully explain the majority of FQHSs observed in the lowest orbital Landau level (LL). However, the ground state of 2DESs at very low LL fillings ($\nu\lesssim1/6$) is still a subject of active discussion ($\nu=nh/eB$ where $n$ is the 2DES density, $h$ is the Planck constant, $e$ is the fundamental charge, and $B$ is the magnetic field). While it was initially suggested theoretically that the transition from the series of FQHSs to a Wigner solid takes place around $\nu\sim1/10$ \cite{Laughlin}, later calculations advocated that it occurs at higher values of $\nu\sim1/6.5$ \cite{Lam,Levesque,Esfarjani,Zhu,Yang}. On the contrary, despite a strong insulating background in the vicinity, hints of a FQHS at $\nu=1/7$ have been reported in the form of an inflection point or a very weak minimum in longitudinal magnetoresistance ($R_{xx}$) measurements of high-quality GaAs 2DESs \cite{Goldman,Pan}.
	
	%In such cases where many-body interaction dominates the energy scale of the system, it is common that different phases of matter are in close competition with each other.

	Recent calculations indicate that in the limit of very low disorder, a FQHS should prevail at $\nu=1/7$ \cite{Zuo}. With the latest breakthrough in the quality of ultra-high-mobility GaAs 2DESs \cite{HighMobility}, it is an opportune moment to revisit this problem and evaluate the behavior of 2DESs in the vicinity of $\nu=1/7$. Here we study the magnetotransport of dilute, ultra-high-quality GaAs 2DESs. All electrical measurements are performed in the van der Pauw geometry on 4 mm $\times$ 4mm square samples, where 8 eutectic InSn contacts are formed on each of the 4 centers of flats and 4 corners. We find a deep minimum in the $R_{xx}$ vs. $B$ trace at $\nu=1/7$, and a weaker but clear sign of a developing minimum is also observed at $\nu=2/13$. Insulating phases, likely deriving from pinned Wigner solid states, still prevail at $\nu\lesssim1/6$, but our data strongly suggest that FQHSs emerge at filling factors $\nu=p/(2mp\pm1)$ even in this very-low-$\nu$ limit. The $\nu$-dependent activation energies ($E_A$) of $R_{xx}$ taken from the relation $R_{xx}\propto e^{E_A/2kT}$ also display local minima at $\nu=1/7$ and 2/13, further corroborating this inference. Compared to previous samples with similar 2DES densities, we also obtain a factor of $\sim2$ to 3 higher values for the $E_A$ of the insulating phases at $\nu\neq p/(2mp\pm1)$. Given that it has been suggested that disorder in samples suppress the $E_A$ of Wigner solid states to lower values \cite{ChaAct1,ChaAct2,JainAct}, the significantly larger $E_A$ values we see are consistent with the fact that our samples are of much higher quality than before.

 \begin{figure*}[t]
\centering
    \includegraphics[width=.95\textwidth]{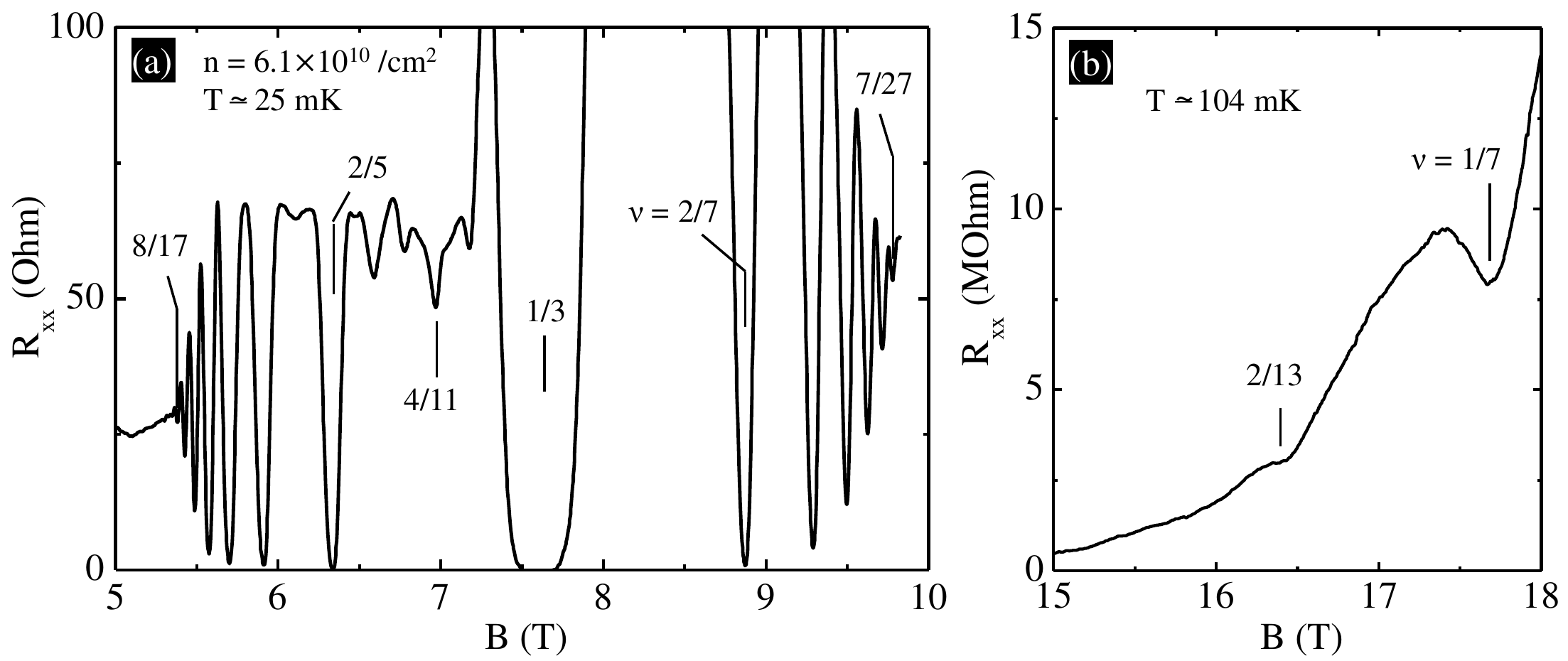}

%\begin{figure*} [t]
%  \begin{center}
%    \psfig{file=Fig1_r10.png, width=0.95\textwidth }
% \end{center}
 \caption{\label{fig1} Representative magnetoresistance ($R_{xx}$) traces of our ultra-high-quality 2DES confined to a 64-nm-wide GaAs quantum well. The 2DES density is $n=6.1\times10^{10}$ /cm$^2$ and the mobility is $\mu=25\times10^6$ cm$^2$/Vs. (a) Low-field trace measured at $T\simeq25$ mK. The magnetic field positions of several LL fillings are marked. The sample quality is extraordinary, demonstrated by the emergence of high-order fractional quantum Hall states up to $\nu=8/17$ near $\nu=1/2$ and $\nu=7/27$ near $\nu=1/4$. (b) High-field trace measured at $T\simeq104$ mK, exhibiting a deep minimum at $\nu=1/7$ and an inflection at $\nu=2/13$.}
\end{figure*}

	Figure 1 shows representative $R_{xx}$ traces of a sample we used in this study. As can be seen in Fig. 1(a), which depicts the low magnetic field data (taken at $T\simeq25$ mK), even though the electron density is only $n=6.1\times10^{10}$ /cm$^2$, high-order FQHSs up to $\nu=8/17$ and 7/27 can be seen near $\nu=1/2$ and 1/4, respectively. The presence of other delicate FQHSs such as the $\nu=4/11$ state is also evident. This is remarkable considering that a sample with $\sim60$\% higher density was required to observe high-order FQHSs such as the $\nu=6/23$ state in previous generation samples typically used to study very low $\nu$ \cite{Pan}. This unequivocally demonstrates that our sample is of much higher quality than before \cite{HighMobility}. The fact that the mobility of our sample is $\sim2.5$ times higher than the sample of Ref. \cite{Pan} despite the lower density corroborates this view. 
	
	This enhanced 2DES quality has a significant impact on $R_{xx}$ when $\nu\lesssim1/6$, as shown in Fig. 1(b). The deep minimum at $B\simeq17.7$ T immediately catches the eye. At the electron density of our sample, this field corresponds to $\nu=1/7$. The apparent contrast of $R_{xx}$ with the insulating background in the vicinity is very suggestive of a FQHS existing at this filling. Also supporting this argument is the fact that a weaker but qualitatively similar feature is observed at $\nu=2/13$, where the next-order FQHS would be expected according to the Jain sequence $\nu=p/(2mp\pm1)$ \cite{JainBook,Jain}. Here $p$ is a positive integer and $2m=6$ is the number of flux quanta attached to the electrons to form composite fermions in this regime. The data we present here greatly strengthen earlier experimental studies that proposed developing FQHSs in the $\nu\lesssim1/6$ regime \cite{Goldman,Pan}.

	It is important to note that the data shown in Fig. 1(b) were taken at $T\simeq104$ mK. Despite the exceptional quality of the sample, we still find that strong insulating phases dominate the magnetotransport in the low-$\nu$ limit. In this regime, $R_{xx}$ values easily exceed several tens of M$\Omega$ when the sample is cooled to the base temperature of our dilution refrigerator ($T\simeq25$ mK), making it difficult to accurately analyze the 2DES. This tendency is qualitatively similar to what was reported in the past \cite{Goldman,Pan}. Still, we were able to probe the interacting behavior of electrons near $\nu=1/6$ in our sample over a reasonable range of temperatures $(86\leq T \leq130$ mK). 
	
	Figure 2, where the $R_{xx}$ data are shown at several temperatures, summarizes these results. For the entire range of magnetic fields shown in Fig. 2, the presence of an insulating background is apparent and becomes more prominent as the temperature is decreased. However, the $R_{xx}$ vs. $B$ profile is not completely monotonic and signatures of FQHSs emerge at different temperatures. These include a deep minimum in $R_{xx}$ at $\nu=1/7$ ($B\simeq17.7$ T), and a less well-developed feature at $\nu=2/13$. 	
	
	The $R_{xx}$ minimum at $\nu=1/7$ appears to be most pronounced when $T\simeq104$ mK. At lower temperatures the insulating background becomes dominant and engulfs the $\nu=1/7$ minimum. At higher temperatures, the background becomes less insulating but the relative depth of the minimum also decreases. Phenomenologically, it appears that an insulating phase and a FQHS coexist at this filling, and the increase in temperature melts the insulating phase and destroys the FQHS simultaneously. This makes it rather difficult to perform standard gap analysis using the expression $R_{xx}\propto e^{-E_g/2kT}$, where $E_g$ is the energy gap of the FQHS and $k$ is the Boltzmann constant. We speculate that this type of behavior may be caused by local density variations over the size of the sample, which is a $4\times4$ mm$^2$ van der Pauw in our case. As the 2DES studied here has a low electron density, even a small amount of density inhomogeneity could significantly alter the local filling factor at a specific magnetic field, especially in the very-low-$\nu$ limit. For example, with a fluctuation on the order of $\Delta n\sim10^9$ /cm$^2$, which is reasonable in samples such as ours \cite{PL}, the magnetic field that corresponds to $\nu=1/7$ varies by $\simeq0.3$ T when $n=6.1\times10^{10}$ /cm$^2$. Since theory suggests that the range of $\nu$ where FQHSs prevail is very narrow for six-flux composite fermion based FQHSs \cite{Zuo}, such density inhomogeneities could locally perturb the ground state of the 2DES and smear features out near $\nu=1/7$. This picture could also explain the insulating, yet FQHS-resembling feature at $\nu=2/13$. 
	
	%As the 2DES studied here has a low electron density, fluctuations even on the order of $\Delta n\sim10^9$ /cm$^2$ could significantly alter the local filling factor at a specific magnetic field, especially in the very-low-$\nu$ limit. Theory suggests that the range of $\nu$ where FQHSs prevail is very narrow for six-flux composite fermion based FQHSs \cite{Zuo}, and such density fluctuations could locally perturb the ground state of the 2DES at $\nu=1/7$. This picture could also explain the insulating, yet FQHS-resembling feature at $\nu=2/13$. 
	
	%A similar trend has been reported in GaAs 2DESs during the early investigations of the $\nu=1/5$ FQHS \cite{Pseudo}. 
	%For example, while $\nu=1/7$ is at $B=17.7$ T when $n=6.1\times10^{10}$ /cm$^2$, it is at $B=17.9$ T when $n=6.2\times10^{10}$ /cm$^2$.

	%The gradual disappearance of the feature in the $R_{xx}$ data as the temperature increases is characteristic of a gapped quantum Hall state, and is consistent with our earlier supposition of a FQHS at this filling. Typically, this behavior should extend to the $T=0$ K limit for a FQHS, which is not the case for the data shown in Fig. 2. At temperatures lower than $T\simeq100$ mK, the $R_{xx}$ minimum at $\nu=1/7$ 
	
	%While weaker, features in the $R_{xx}$ data at fillings $\nu=2/13$ and 3/17 also resemble FQHSs competing with the strong insulating background.

\begin{figure}[t]
 
 \centering
    \includegraphics[width=.45\textwidth]{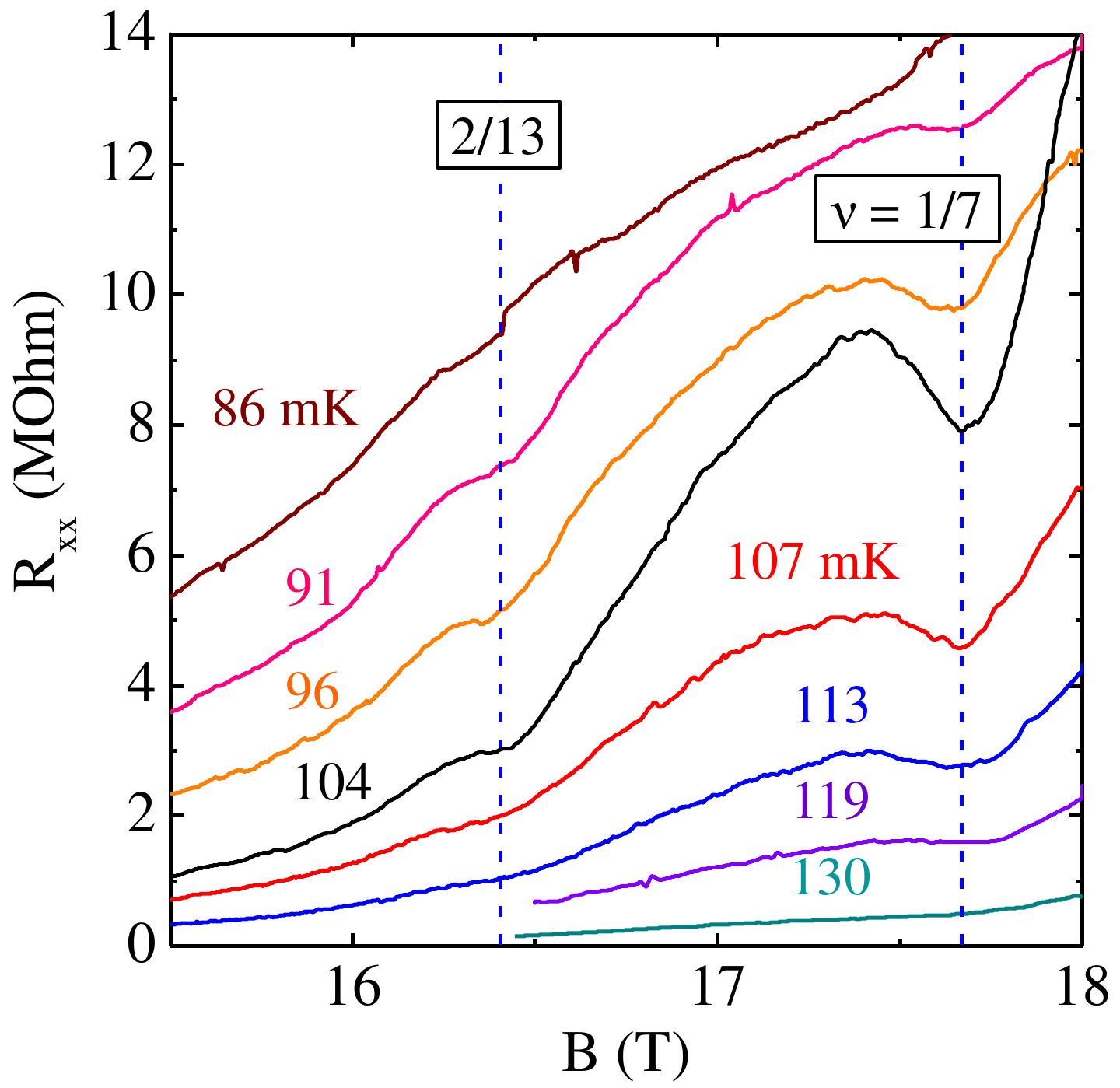} 
% \begin{center}
%    \psfig{file=Fig2_r9.png, width=0.45\textwidth }
%  \end{center}
  \caption{\label{fig2} $R_{xx}$ vs. $B$ traces taken at low LL fillings ($\nu\lesssim1/6$) at various temperatures. Magnetic field positions for $\nu=1/7$ and 2/13 are marked by the blue dashed lines. There is a deep minimum developing in $R_{xx}$ at $\nu=1/7$ for temperatures near $T\simeq104$ mK, but the minimum vanishes into the highly insulating background at lower and higher temperatures. Hints of an emerging FQHS can also be seen at $\nu=2/13$.}
\end{figure}

\begin{figure}[t]

\centering
    \includegraphics[width=.45\textwidth]{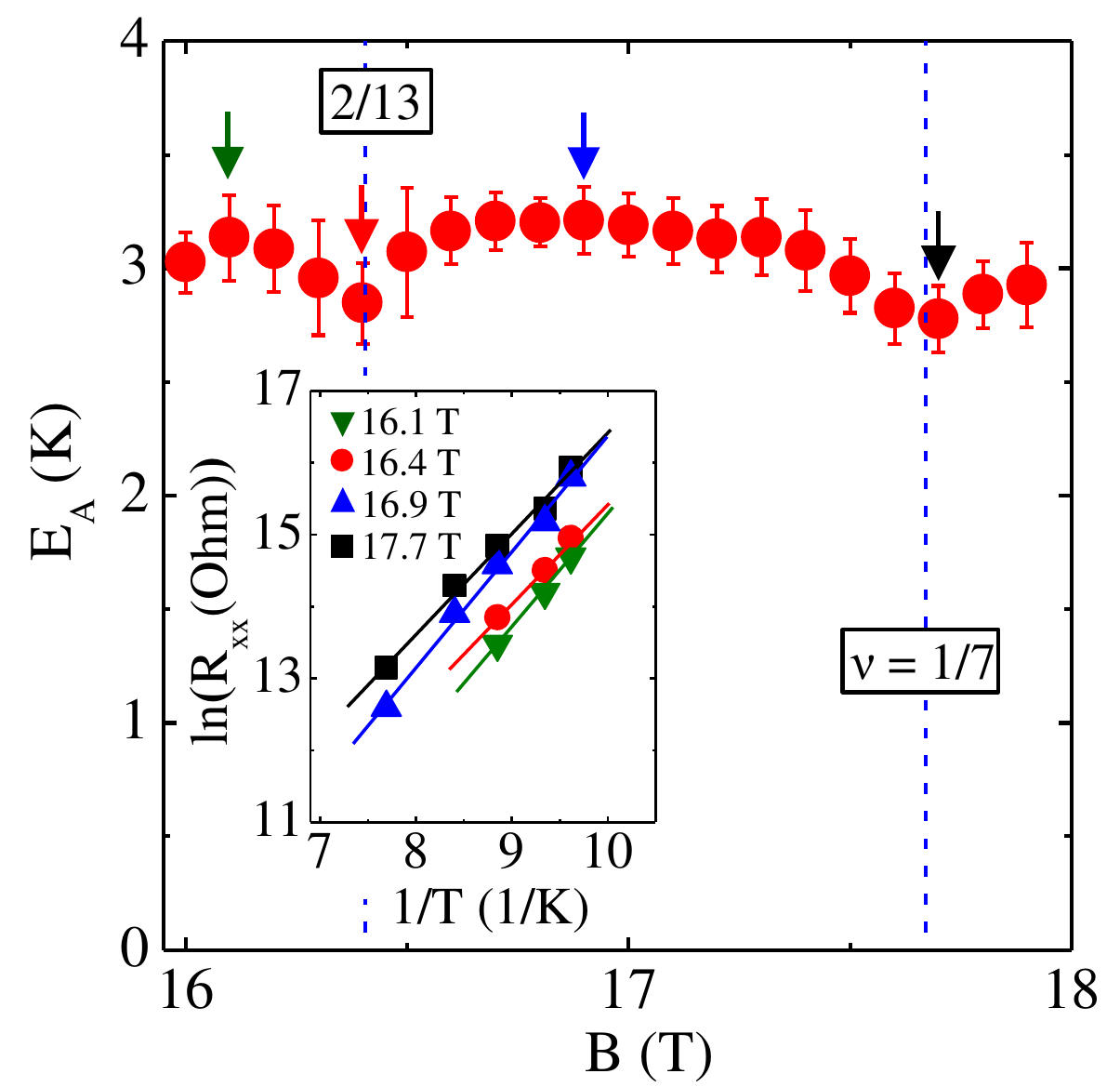} 
%  \begin{center}
 %   \psfig{file=Fig3_r9.png, width=0.45\textwidth }
 % \end{center}
  \caption{\label{fig3} The activation energy ($E_A$) for the $R_{xx}$ data shown in Fig. 2, deduced from the relation $R_{xx}\propto e^{E_A/2kT}$. Magnetic field positions for $\nu=1/7$ and 2/13 are marked with blue dashed lines. Data sets plotted in the inset show $ln$($R_{xx}$) vs. $1/T$ at various magnetic fields, each color coded according to the magnetic field values marked by the arrows in the main figure. The solid lines drawn in the inset show the linear fits used to obtain the $E_A$ values. The error bars in the main figure correspond to the standard deviation in the slope of these linear fits.}
\end{figure} 

%slope of the linear fits for each set of activation data, drawn using solid lines, correspond to 

	The origin of the strong insulating phases observed at very-low $\nu$ warrants some discussion. We estimate that the impurity concentration is on the order of $\sim1\times10^{13}$ /cm$^3$ in our sample \cite{HighMobility}, which implies an average inter-impurity spacing of $\sim0.5$ $\mu$m. This is significantly larger than the average inter-electron spacing of $\sim0.04$ $\mu$m computed from the 2DES density of $n=6.1\times10^{10}$ /cm$^2$. Given this situation, it seems likely that the low-temperature insulating behavior we see in Figs. 1(b) and 2 derives from the formation of Wigner solid phases rather than simple carrier localization.  
	
%Given the extremely low amount of disorder in our 2DES, it is likely that the insulating phases observed at low Landau level fillings derive from Wigner solid formation.  

	A useful method to quantitatively analyze such many-body states is to evaluate their temperature activated characteristics in magnetotransport. The energy scale $E_A$ obtained from the relation $R_{xx}\propto e^{E_A/2kT}$ is commonly associated with the melting/defect formation temperature of the Wigner solid \cite{Wigner2,Wigner3,ChaAct1,ChaAct2,JainAct,WillettAct,ChuiAct,JiangAct2,PaalanenAct,DuAct}. Figure 3 shows the activation energy of $R_{xx}$ in our sample at magnetic fields where $\nu\lesssim1/6$. The inset shows the $R_{xx}$ vs. $1/T$ plots used to determine the $E_A$ values shown in the main figure for several magnetic fields \cite{footnote}. It is clear from the data in Fig. 3 that there are $E_A$ minima emerging at $\nu=1/7$ and 2/13 where FQHSs are suspected to exist from the $R_{xx}$ vs. $B$ data in Fig. 2. 
	
	Local minima in $E_A$ have been interpreted as precursors for developing FQHSs in the past at higher LL fillings such as $\nu=1/5$ \cite{Wigner2,Wigner3,WillettAct,JiangAct2,PaalanenAct,DuAct}. For example, much like the data for $\nu=1/7$ in Fig. 2, early measurements of the $\nu=1/5$ FQHS only displayed a reasonably deep minimum surrounded by an insulating background in the $R_{xx}$ at $\nu=1/5$. Further analysis of the filling factor dependent $E_A$ revealed an inflection point at $\nu=1/5$, which was taken to be an indicator of a developing FQHS \cite{WillettAct}. Indeed, as sample quality improved, the $R_{xx}$ minimum at $\nu=1/5$ became more prominent and approached near-zero values. Concurrently, the $\nu$ vs. $E_A$ plots in this regime unveiled a dome-like structure reminiscent of a reentrant phase transition on the flanks of $\nu=1/5$ \cite{Wigner2,Wigner3,WillettAct,JiangAct2,PaalanenAct,DuAct}. It is entirely conceivable that samples with even higher quality and density uniformity than our present samples would exhibit a FQHS with vanishing $R_{xx}$ (as $T\rightarrow0$) at $\nu=1/7$. 
	
	At magnetic fields other than those corresponding to $\nu=1/7$ and 2/13, the activation energy in Fig. 3 is fairly constant, typically showing values around $E_A\simeq3.0$ K. A qualitatively similar trend has been observed for $E_A$ of insulating phases near $\nu\lesssim1/5$ in previous samples with comparable electron density, but with much lower quality and mobility \cite{JiangAct2,PaalanenAct}. The data yielded smaller values of $E_A\simeq1.5$ K. Following theoretical models that link the activation energy with the defect formation energy of a Wigner solid, the larger $E_A$ in our sample is consistent with the expectation that lower disorder in the 2DES should yield larger $E_A$ values \cite{ChaAct1,ChaAct2,JainAct}. In fact, the $E_A$ values we observe are remarkably close to the defect formation energies computed for composite fermion-based crystals. For example, at $B=16.9$ T ($\nu\simeq0.15$), theory predicts a defect formation energy between 2.5 to 4.2 K depending on the number of vortices attached to the defect \cite{JainAct}, while in our experiments we deduce $E_A\simeq3.2$ K. The relative invariance of $E_A$ we see in our sample over the range of measured $\nu$ also agrees well with these calculations \cite{JainAct}. 
	
\begin{figure}[t]

\centering
    \includegraphics[width=.45\textwidth]{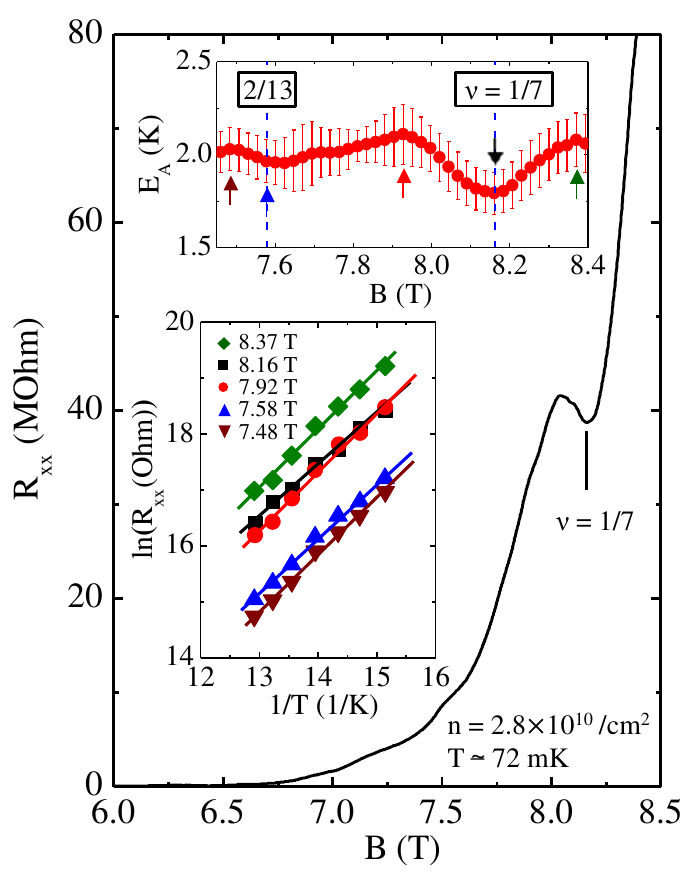} 
%  \begin{center}
 %   \psfig{file=Fig3_r9.png, width=0.45\textwidth }
 % \end{center}
  \caption{\label{fig4} Signatures of a $\nu=1/7$ FQHS emerging in another ultra-high-quality GaAs sample with a lower electron density of $n=2.8\times10^{10}$ /cm$^2$ and mobility $\mu=15\times10^6$ cm$^2$/Vs. The main figure is a high-field $R_{xx}$ trace of the sample taken at $T=72$ mK, which exhibits a deep minimum at $\nu=1/7$. Top inset shows magnetic field dependent $E_A$ values, and some representative $ln$($R_{xx}$) vs. $1/T$ data sets used to obtain these results are plotted in the bottom left inset. The $ln$($R_{xx}$) vs. $1/T$ data sets are color coded according to magnetic field, marked by the arrows in the top inset.}
\end{figure} 
	
	The characteristics we measure for many-body electron phases at $\nu\lesssim1/6$ do not appear to be sample or density specific. As summarized in Fig. 4, in another ultra-high-quality GaAs 2DES with a lower density of $n=2.8\times10^{10}$ /cm$^2$ and mobility $\mu=15\times10^6$ cm$^2$/Vs, we see data that closely resemble those of Figs. 1-3. As seen in the main part of Fig. 4, there is a deep minimum in the $R_{xx}$ vs. $B$ trace of this lower density sample at $\nu=1/7$ when $T\simeq72$ mK, which is the temperature at which the minimum is most pronounced. Analogous to the higher-density sample, at lower temperatures all features are devoured by the insulating background, and at higher temperatures both the minimum and insulating background weaken in strength. The top inset of Fig. 4 shows $E_A$ vs. $\nu$ for this sample near $\nu=1/7$, while the lower inset presents the activation data. Consistent with Fig. 3 data, $E_A$ is mostly constant, with minima showing at magnetic fields corresponding to $\nu=1/7$ and 2/13. One quantitative difference is that the $E_A$ values are generally smaller ($E_A\simeq2.0$ K) for the lower-density sample. This is reasonable considering that the defect formation energy of Wigner solids at $\nu\lesssim1/6$ is expected to scale inversely with the magnetic length, $\ell_B=\sqrt{\hbar/eB}$ \cite{JainAct}. Furthermore, the lower electron density implies less screening from impurities, which could also reduce the defect formation energy \cite{JainAct}. 
	
	While the features observed in both the $R_{xx}$ and $E_A$ vs. $B$ plots for both samples clearly demonstrate that the improvement in sample quality allows a better visualization of the $\nu=1/7$ and its satellite FQHSs, it is evident that these states are still in close competition with the background insulating phase. For the insulating phases near $\nu=1/5$, an improvement in the mobility from $\mu=1.7\times10^6$ cm$^2$/Vs to $\mu=2.0\times10^6$ cm$^2$/Vs was enough for 2DESs with densities $n\simeq5\times10^{10}$ /cm$^2$ to show a qualitative difference in the activation energy profiles \cite{WillettAct,JiangAct2}. The lower-mobility sample merely displayed an inflection point at $\nu=1/5$ in the $E_A$ vs. $\nu$ plot \cite{WillettAct}, while for the higher-mobility sample clear signs of reentrant Wigner solid phases were observed on the flanks of a $\nu=1/5$ FQHS (with vanishing $R_{xx}$ as $T\rightarrow0$) \cite{JiangAct2}.  
	
	The data in Fig. 3 are then somewhat surprising given that our $n=6.1\times10^{10}$ /cm$^2$ sample has roughly 3 times fewer impurities compared to the previous sample used to study correlated states near $\nu=1/7$ \cite{Pan,HighMobility}. However, it is important to remember that the $\nu=1/m$ FQHSs become more fragile as $m$ increases. For example, the experimentally measured energy gap for the $\nu=1/3$ FQHS in a high-quality GaAs 2DES is $\simeq8$ K when it is evaluated at $B\simeq15$ T \cite{DuThird}, but the gap for the $\nu=1/5$ FQHS in a similar sample is only $\simeq1$ K even when evaluated at a higher magnetic field of $B\simeq20$ T \cite{Wigner2}. 
	
	Theoretical studies support these findings, typically citing gap energies of $\simeq0.1$, 0.03, and 0.007 in units of $e^2/4\pi\epsilon\epsilon_0\ell_B$ for the $\nu=1/3$, 1/5, and 1/7 FQHSs, respectively \cite{Zuo}; here $\epsilon$ is the host material's dielectric constant. Note that these are gaps expected for ``ideal" 2DESs; in realistic samples, the finite thickness of the electron layer, the mixing between the LLs, and disorder would lead to smaller gaps \cite{KAVR}. For our high-density sample in Figs. 1-3, we then expect a maximum theoretical gap of $\simeq1.5$ K for the $\nu=1/7$ FQHS. This is comparable to the value of $\simeq1$ K that has been reported for the phenomenological disorder parameter $\Gamma$, often linked to LL broadening, in state-of-the-art GaAs 2DESs \cite{KAVR}. It is therefore possible that even purer samples are necessary to observe a fully-developed $\nu=1/7$ FQHS. Indeed, theory argues that, although the $\nu=1/7$ FQHS should prevail in the limit of no disorder and perfectly uniform 2DES density, even a very small perturbation to the system would cause insulating crystals and incompressible FQH liquids to coexist \cite{Zuo}, as suggested by our data.
		
	In closing, we comment that the study of correlated-electron phases at high magnetic fields in ultra-low-disorder GaAs samples is especially intriguing because electrons are condensed into a flatband, namely an LL, where the Coulomb energy dominates. For several decades, this has been the platform of choice to investigate many-body electron physics and the competition between various exotic phases such as the FQH liquid and metallic (e.g., composite fermion) phases, and broken-symmetry phases such as Wigner crystals and stripe phases. Recently, it was proposed that flatbands can also emerge in moir\'e-like super-lattice structures in multilayer 2D materials \cite{Macdonald}, and an explosive growth has occurred in this area of research. A particular highlight is the observation of correlated superconducting and insulating phases in magic-angle twisted bilayer graphene \cite{Cao1,Cao2}. It has been pointed out that the knowledge acquired in LL systems can indeed be helpful for understanding the intricate physics of these newer, moir\'e-based, flatband 2D systems \cite{Tarnopolsky,Balents}. Our data, taken in the highest-quality (lowest-disorder) 2DESs, provide valuable information and incentive for future studies in other flatband systems with improved quality and comparable settings.

\begin{acknowledgments}
We acknowledge support through the NSF (Grants DMR 2104771 and ECCS 1906253) for measurements, the Department of Energy (DOE) Basic Energy Sciences (Grant DE-FG02-00-ER45841) for sample characterization, and the NSF (Grant MRSEC DMR 1420541), and the Gordon and Betty Moore Foundation (Grant GBMF9615 to L. N. P.) for sample fabrication. A portion of this work was performed at the National High Magnetic Field Laboratory (NHMFL), which is supported by the National Science Foundation Cooperative Agreement No. DMR-1644779 and the state of Florida. We thank S. Hannahs, T. Murphy, A. Bangura, G. Jones, and E. Green at NHMFL for technical support. We also thank J. K. Jain for illuminating discussions.

 \end{acknowledgments}
 
 The data that support the findings of this study are available from the corresponding author upon reasonable request.


\begin{thebibliography}{99}
\bibitem{footnote} The number of data points used to deduce the activation energy in Fig. 3 is limited because of the time constraints at the National High Magnetic Field laboratory. A qualitatively similar behavior is observed in activation energy values deduced for a lower-density sample, presented in Fig. 4. Much more data are available for this sample because it was measured over a longer period in a dilution refrigerator at Princeton.
\end{thebibliography}
\end{document}